\documentclass[aps,pre,preprint,groupedaddress,showpacs]{revtex4-1}

\usepackage{graphicx}

\newcommand{\Ei}{{\rm Ei}}

\begin{document}

\title{Phase transitions in simplified models with long-range interactions}
\author{T.~M.~Rocha Filho, M.~A.~Amato, B.~A.~Mello and A.~Figueiredo}
\affiliation{Instituto de F\'\i{}sica and International Center for Condensed Matter Physics\\ Universidade de
Bras\'\i{}lia, CP: 04455, 70919-970 - Bras\'\i{}lia, Brazil}
\date{\today}
\begin{abstract}
We study the origin of phase transitions in some simplified models with long range interactions. For the self-gravitating
ring model, we were unable to observe a possible new phase transition predicted in a recent paper by Nardini and Casetti
from an energy landscape analysis. Instead of it we observed a sharp, although
without any non-analyticity, change from a core-halo to a core only configuration in the spatial distribution functions
for low energies. By introducing a new class of solvable simplified models without any critical points in the potential energy,
we show that a behaviour similar to the thermodynamics of the ring model is obtained, with a first order phase transition from an almost
homogeneous high energy phase to a clustered phase, and the same core-halo to core configuration transition at lower energies.
We discuss the origin of these features for the simplified models, and show that the first order phase transition comes
from the maximization of the entropy of the system as a function of energy and an order parameter, as previously discussed by Hahn and Kastner,
which seems to be the main mechanism causing phase transitions in long-range interacting systems.
\end{abstract}

\maketitle

\section{Introduction}

The Statistical Mechanics of systems with long-range interactions
presents many challenging problems, well illustrated by the
existence of quasi-stationary non-Gaussian states with diverging life-times as the number of particles increases,
negative microcanonical heat capacity, inequivalence of ensembles and non-ergodicity~\cite{prrev,proc1,proc2,proc3,nos,eplnos}. Examples of
systems with long-range forces comprise self-gravitating systems~\cite{padma,levin0}, non-neutral
plasmas~\cite{levin,levin2}, two-dimensional flows~\cite{twodflows} and some models such as the
free electron laser~\cite{fel}, wave-particle interaction in plasmas~\cite{buyl} and the
self-gravitating ring model~\cite{ring1,ring2}. An
interaction potential is long ranged if it decays for long distances as $r^{-\nu}$,
with $\nu<d$, and $d$ the spatial dimension.
Since different statistical ensembles may not be equivalent for long-range forces, the choice
of the working ensemble must be dictated by the physical constraints on the system, and the microcanonical ensemble
is the fundamental ensemble in the sense that information for other ensembles can be obtained
from the knowledge of the microcanonical entropy, the converse being not always true~\cite{touchette}.
As a consequence the nature of phase transitions may also depend on the physical constraints.
It is interesting to note that phase transitions occur in one-dimensional classical systems with short-range interactions
provided some conditions are satisfied~\cite{ruelle,cuesta1,cuesta2},
while for long-range interactions they are more commonly observed~\cite{ring1,ring2,hmf}.

The study of self-gravitating systems with a large number of particles (as in globular clusters or in galaxies)
is a very difficult task, and the determination of at least some of its properties from statistical mechanics
is one direction that have been explored during the last decades (see~\cite{padma2} and references therein).
Nevertheless some problems are still open and simplified models have been introduced to clarify those
issues such as the determination of the distribution function after the process of violent relaxation~\cite{lyndenbell,arad,levin3},
the dynamics of gravothermal collapse~\cite{sire}, and the nature of phase transitions~\cite{chavanis1}.
It is known that no global maximum for the entropy of a self-gravitating system exists unless it is enclosed in
a container (usually an spherical one with a given large radius), and the $1/r$ potential is somehow bounded from below by a cutoff,
by the introduction of a small softening parameter $\epsilon$ leading to a potential in $1/(r+\epsilon)$,
or by considering the Pauli exclusion principle for self-gravitating fermions~\cite{padma,padma2}.
Examples of simplified models include the self-gravitational
sheet model~\cite{valegas}, the ring model~\cite{ring1} and the Hamiltonian Mean Field (HMF) model~\cite{hmf}, all one-dimensional,
and two-dimensional gravity~\cite{padma,levin0}.

In this paper we are interested in discussing the causes of phase transitions in simplified models with long-range interactions.
Different approaches are reported in the literature to study equilibrium phase transitions in those systems~\cite{prrev,chavanis1,largedev,metodo}.
The energy landscape method have been used to predict phase transitions in systems with short-range interactions,
and in some solvable models with long range interactions, based on the topological approach
to phase transitions~\cite{casetti1,kastner1,a1}. This approach can be briefly summarized as follows.
For a classical system described by a Hamiltonian of the form
\begin{equation}
H(q,p)=\sum_{i=1}^N\frac{p_i^2}{2m}+V(q),
\end{equation}
where $N$ is the number of particles and $V(q)$ is the interaction potential,
the topological approach relies on the study the topology changes 
of the following manifolds in the configuration space $\Gamma_N$:
\begin{equation}
{\cal M}_v=\{q\in\Gamma_N|V(q_1,\ldots,q_N)\le Nv\},
\end{equation}
and
\begin{equation}
\Sigma_v=\{q\in\Gamma_N|V(q_1,\ldots,q_N)=Nv\},
\end{equation}
with  $v$ a fixed value for the potential energy per particle.
For most systems, both manifolds lead to the same thermodynamical properties~\cite{ruelle}. This is not necessarily the case for long-range interacting systems
(see discussion in section III).
The study proceeds using Morse theory, where a smooth potential energy is a Morse function (see~\cite{casetti1,kastner1,a1,milnor,nicolaiescu} for details).
For short-range confining potential,
Franzosi and Pettini presented a proof for a theorem ensuring that a phase transition is always associated to a topology
transition in the manifold ${\cal M}_v$ at $V=V_c=Nv_c$ the critical value of the potential at the phase
transition~\cite{franzoni,a2,a3}. In fact, there is a one-to-one relation between topology changes and critical points $\overline{q}$ of $V(q)$,
such that $\nabla_{q_i} V|_{q=\overline{q}}=0$, $i=1,\ldots,N$, provided $V$ is a proper Morse function~\cite{kastner2}.
Such critical points are also called saddle points in the literature on energy landscapes.

On the other hand, not all critical points
are associated to phase transitions, and more recently a new necessary condition, the so-called Kastner-Schnetz-Schreiber (KSS) criterion.
was formulated stating that only asymptotically flat critical points can lead to a phase transition in the thermodynamic limit
(see~\cite{kastner2} for more details).
The microcanonical entropy per particle (or degree of freedom) $s(e)=S(E)/N$, $e=E/N$,
can be non-analytic in the thermodynamic limit at a given energy $E=E_c$
if: (i) there is a sequence $\{\overline{q}^{(N)}\}_{N=1}^\infty$ of critical points of $N$-particle potential $V$ such that
\begin{equation}
\lim_{N\rightarrow\infty} V(\overline{q}^{(N)})=Nv_c,
\label{critcond1}
\end{equation}
where $v_c=\langle V \rangle_{E=E_c}/N$ is the statistical average value of the potential energy per particle at the transition, and
(ii) the Hessian matrix of $V$ at the critical points $\overline{q}^{(N)}$, given by
\begin{equation}
{\bf H}_V^{(N)}=\left.\left(\frac{\partial^2 V}{\partial q_i\partial q_j}\right)\right|_{q=\overline{q}^{(N)}},
\label{hessian}
\end{equation}
satisfies
\begin{equation}
\lim_{N\rightarrow\infty} \left|{\rm Det} \left({\bf H}_V^{(N)}\right)\right| ^{1/N}=0.
\label{critcond2}
\end{equation}
Equation~(\ref{critcond2}) expresses the asymptotic flatness condition for the critical point.
This approach was applied with success to the mean field $XY$ and $k$-trigonometric models~\cite{kastner2,kastner3}.

Some systems nevertheless are known to display phase transitions not related to a topology change in the manifold
${\cal M}_v$, and particularly in some cases with long-range interacting potentials~\cite{kastner1}.
On the other hand, the KSS criterion was recently applied to the
self-gravitating ring model, indicating the possibility of a new phase transition
hitherto undetected~\cite{ring3}. By going beyond previous studies of this model~\cite{ring1,ring2},
we were unable to observe the predicted new phase transition. Instead, we observed a
structural transformation at lower energies from a core-halo to a core only configuration, such that the halo rapidly drops
to almost zero. It is reminiscent of the structural change discussed by Aronson and Hansen for an isothermal
sphere with a short distance cutoff~\cite{aronson} (see also~\cite{padma}).
In order to further discuss these issues, we introduce a new class of
solvable simplified models with attractive long-range forces with the same qualitative behaviour as the ring model.
We also discuss for long-range interacting systems admitting a mean-field description a possible common origin of phase-transitions.

The structure of the paper is as follows: in section~II we discuss the thermodynamics of the ring model analyzing
the conjecture by Nardini and Casetti on the possible existence of an additional phase transition. In section~III
we introduce a class of solvable models, and discuss with more detail a representative system of this class. We close
the paper with a discussion of our main results and some concluding remarks in section~IV.

\section{Thermodynamics of the ring model}

The ring model consists of a system with $N$ particles of mass $m$ constrained to a circular ring of radius $R$,
and interacting by their mutual gravitational potential. By a judicious choice of units, the Hamiltonian
for the model is written as~\cite{ring1}:
\begin{equation}
H=\frac{1}{2}\sum_{i=1}^N p_i^2+\frac{1}{2N}\sum_{i,j=1}^N V_{ij},
\label{hamgen}
\end{equation}
where the pair interaction potential:
\begin{equation}
V_{ij}=-\frac{1}{\sqrt{2}\sqrt{1-\cos(\theta_i-\theta_j)+\epsilon}},
\label{pairpot1}
\end{equation}
with $\epsilon>0$ the softening parameter, introduced in numerical simulations of self-gravitating systems in order to avoid
the divergence of the potential for $r\rightarrow 0$~\cite{aarseth}. The coordinate $\theta_i\in[-\pi,\pi)$ stands for the angle
position of the $i$-th particle on the circle and the $1/N$ Kac prefactor in the second terms in the right hand side of eq.~(\ref{hamgen})
is introduced so that the total energy of the system is extensive, albeit not additive, and can be achieved by a choice of time units.
The minimum possible energy per particle is
$e_0=-1/(2\sqrt{2\epsilon})$, and in this work we rescale the energy unit such that the minimal energy per particle is $-1$ for all
values of $\epsilon$. This results in the pair interaction potential:
\begin{equation}
V_{ij}=V(\theta_i-\theta_j)=-\frac{2\sqrt{\epsilon}}{\sqrt{1-\cos(\theta_i-\theta_j)+\epsilon}}.
\label{pairpot2}
\end{equation}

This model is known to have a phase transition from a homogeneous phase at higher energies to an inhomogeneous phase at lower energies.
This transition can be continuous or discontinuous for higher or lower values of $\epsilon$ respectively.
For smaller values of $\epsilon$ the system also has an energy range with a negative heat capacity.
These results can be obtained by numerical simulations~\cite{ring1} or a variational method~\cite{ring2} in the microcanonical ensemble.
Microcanonical Monte Carlo simulations can also be used and will be explored in this work together with a variational method.
In Reference~\cite{ring3} Nardini and Casetti applied the energy landscape method by identifying different kinds of critical (saddle) points of the
potential (fixing $\theta_1=0$ such that the fixed points are isolated):
$0-\pi$ critical points with $N_\pi$ particle at $\theta=\pi$ and $N-N_\pi$ particles at $\theta=0$, 
and polygonal critical points with $N/p$ particles in the $p$ vertices of a regular polygon.
They shown that the sequence of polygonal critical points satisfy the KSS criterion for the
value of $v_c=V/N$ corresponding to the transition from the homogeneous to inhomogeneous phases, and
also that $0-\pi$ critical points satisfy  the KSS criterion for critical values of the potential per particle given by:
\begin{equation}
v_c=-\frac{8+2\epsilon[6+\epsilon(5+2\epsilon)]}{[(2+\epsilon)^{3/2}+\epsilon^{3/2}]^2},
\label{vctrans}
\end{equation}
where this expression is already written in units of $|e_0|$. For almost all values of $\epsilon$ the transition occurs
at very low energies (near $e=-1$), except for $\epsilon$  close to $1$. Therefore these authors
conjectured on the existence of a previously unobserved phase at lower energies, with the fraction of particles crossing at $\theta=\pi$ for
$v<v_c$ vanishing (see~\cite{ring3} for details). In Reference~\cite{casetti2} these same authors introduced
a simplified model of self-gravitating particles derived from approximations on the ring model.
They found that particles form a very dense cluster, or core, surrounded by particles in a cloud, or halo structure, moving almost freely.
For very small energies the fraction of particles in the halo is very small, but never vanishes.
The model in~\cite{casetti2} does indeed reproduce quite closely the thermodynamics of the ring model
and has many similarities with the Thirring model~\cite{thirring}. In fact, as we argue below, both the ring model and the simplified models
in section~IV show the same qualitative behaviour expected to occur for attractive long-range and regularized potentials.

\begin{figure}[ptb]
\begin{center}
\scalebox{0.3}{{\includegraphics{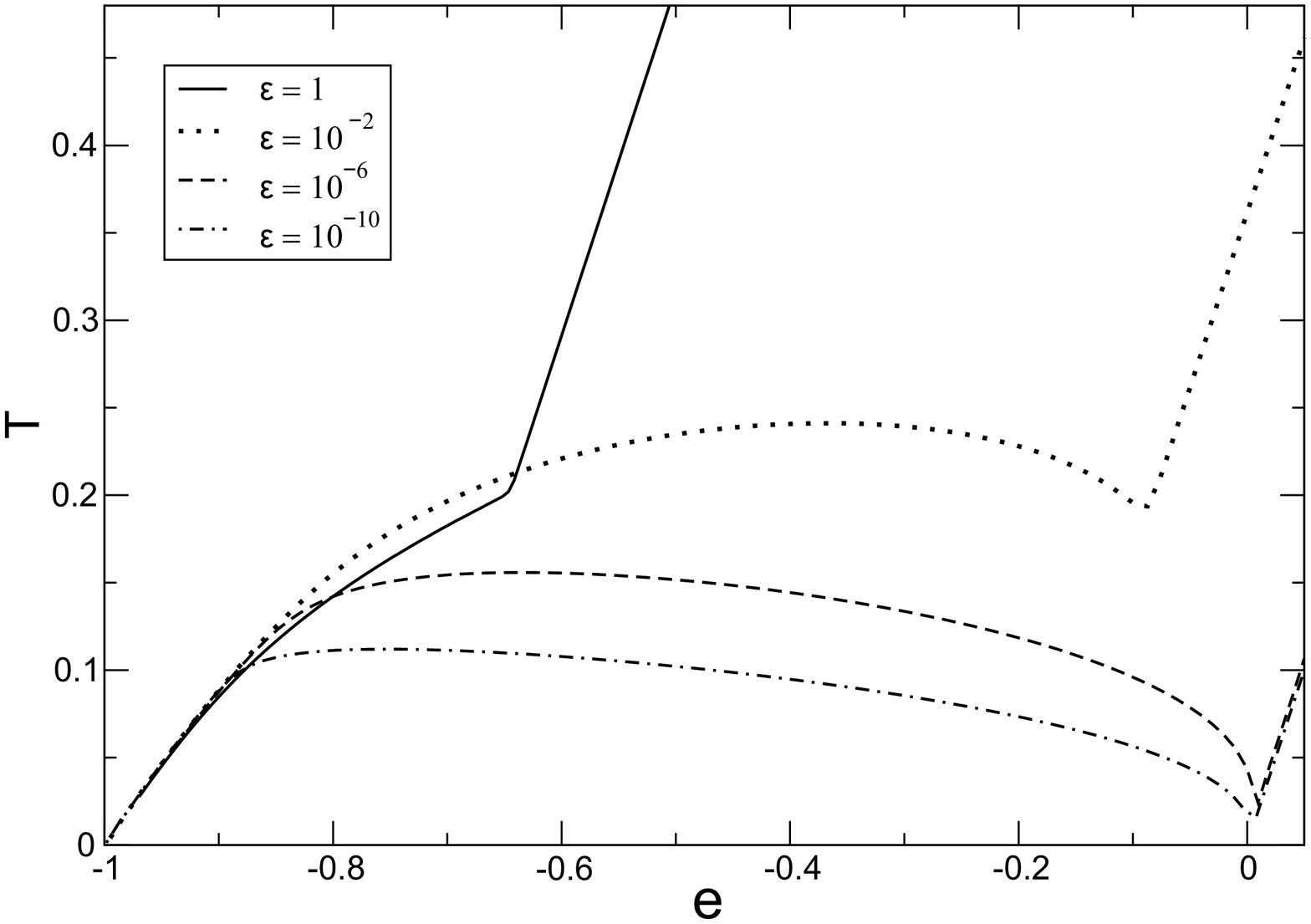}}}
\scalebox{0.3}{{\includegraphics{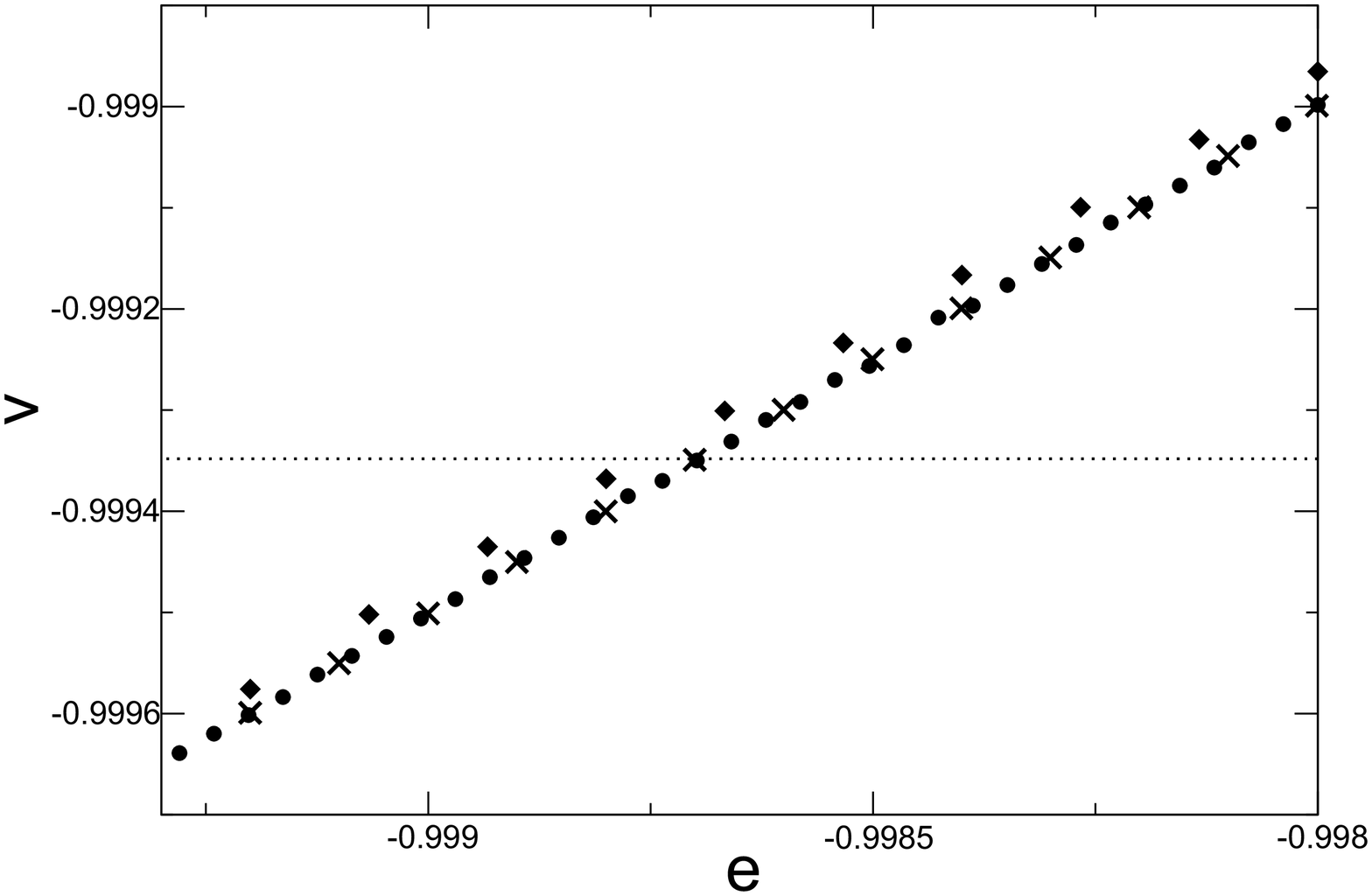}}}
\end{center}
\caption{Left panel: Caloric curves for the potential in~(\ref{pairpot2}) for some values of $\epsilon$.
Right panel: Potential energy as a function of total energy (per particle) for $\epsilon=10^{-2}$ on an energy interval comprising
the value $v_c\approx-.99935$ (dotted line) from the EM (diamonds), MC (circles) and MD with $N=100$ (crosses) methods.}
\label{fig1}
\end{figure}

\begin{figure}[ptb]
\begin{center}
\scalebox{0.3}{{\includegraphics{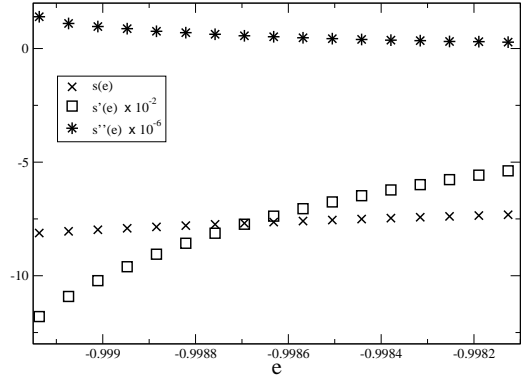}}}
\end{center}
\caption{Entropy and its first and second derivatives with respect to energy around the expected phase-transition in eq.~\ref{vctrans}. The
first derivative $s^\prime(e)$ is divided by 100 and the second derivative $s^{\prime\prime}(e)$ by $10^6$ such that they fit in the same graphic.}
\label{nova1}
\end{figure}

Figure~\ref{fig1} shows the caloric curves
for some values of $\epsilon$ obtained using the Entropy Maximization (EM) method described in Reference~\cite{ring2}.
In this method the distribution is represented in a suitable numerical grid and is determined iteratively, ensuring convergence of the entropy to its maximum for a given energy
in the $N\rightarrow\infty$ limit, within a mean-field description which is exact in this limit.
The same curves can be obtained from a microcanonical Monte Carlo (MC) as described in~\cite{ray} or Molecular Dynamics (MD) simulations
using the fourth-order symplectic integrator in~\cite{yoshida}, as shown in the right panel of figure~\ref{fig1}.
From our results obtained using MC and MD simulations, and the EM method, as shown in figure~\ref{fig1},
no sign of a phase transition is observed at the predicted critical value $v_c\approx-.99935$.
Figure~\ref{nova1} shows the entropy and its first and second derivatives as a function of energy obtained using the EM method, for an energy interval comprising the phase
transition in eq.~(\ref{vctrans}), and no signature of a phase transition is observed.

\begin{figure}[ptb]
\begin{center}
\scalebox{0.3}{{\includegraphics{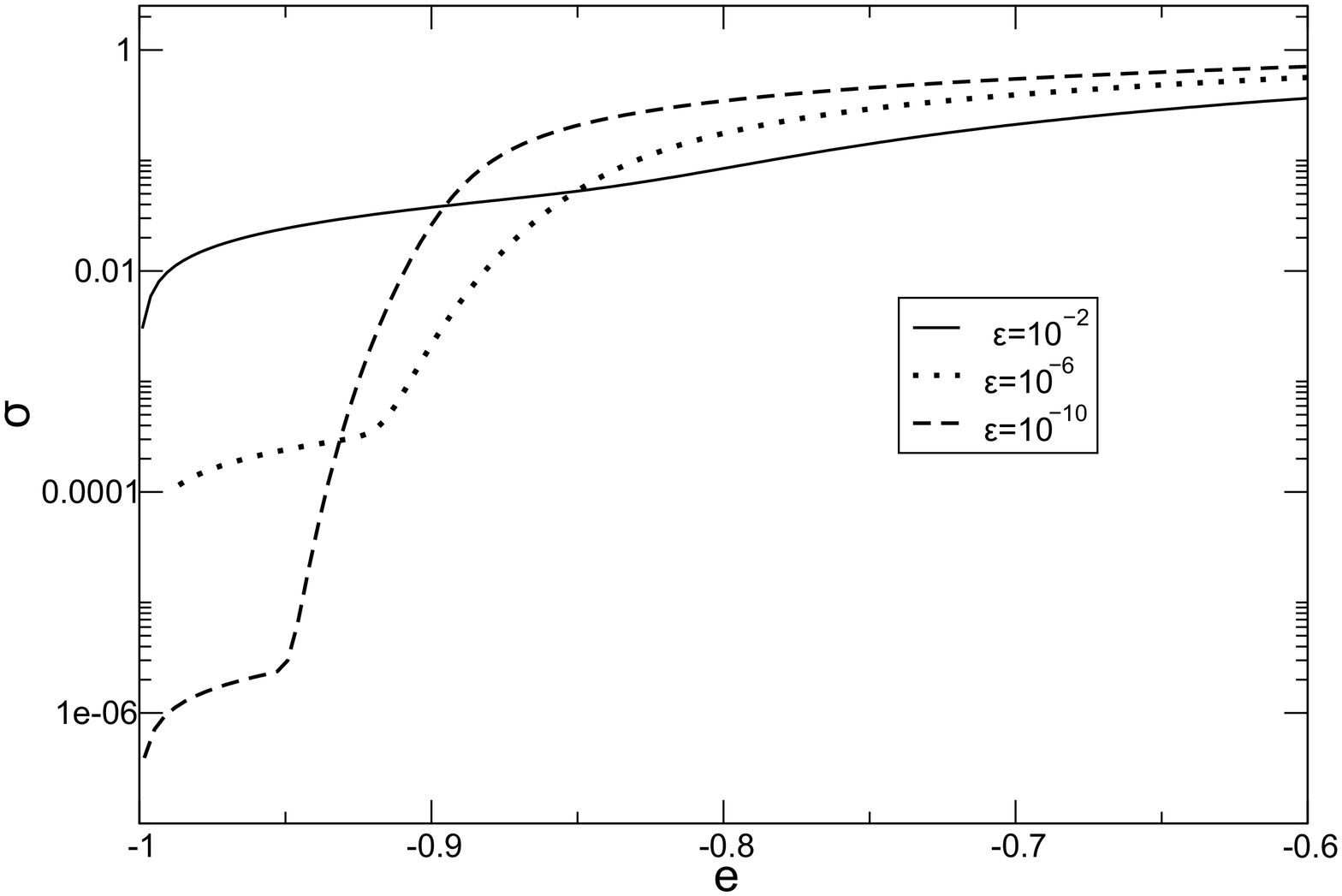}}}
\scalebox{0.3}{{\includegraphics{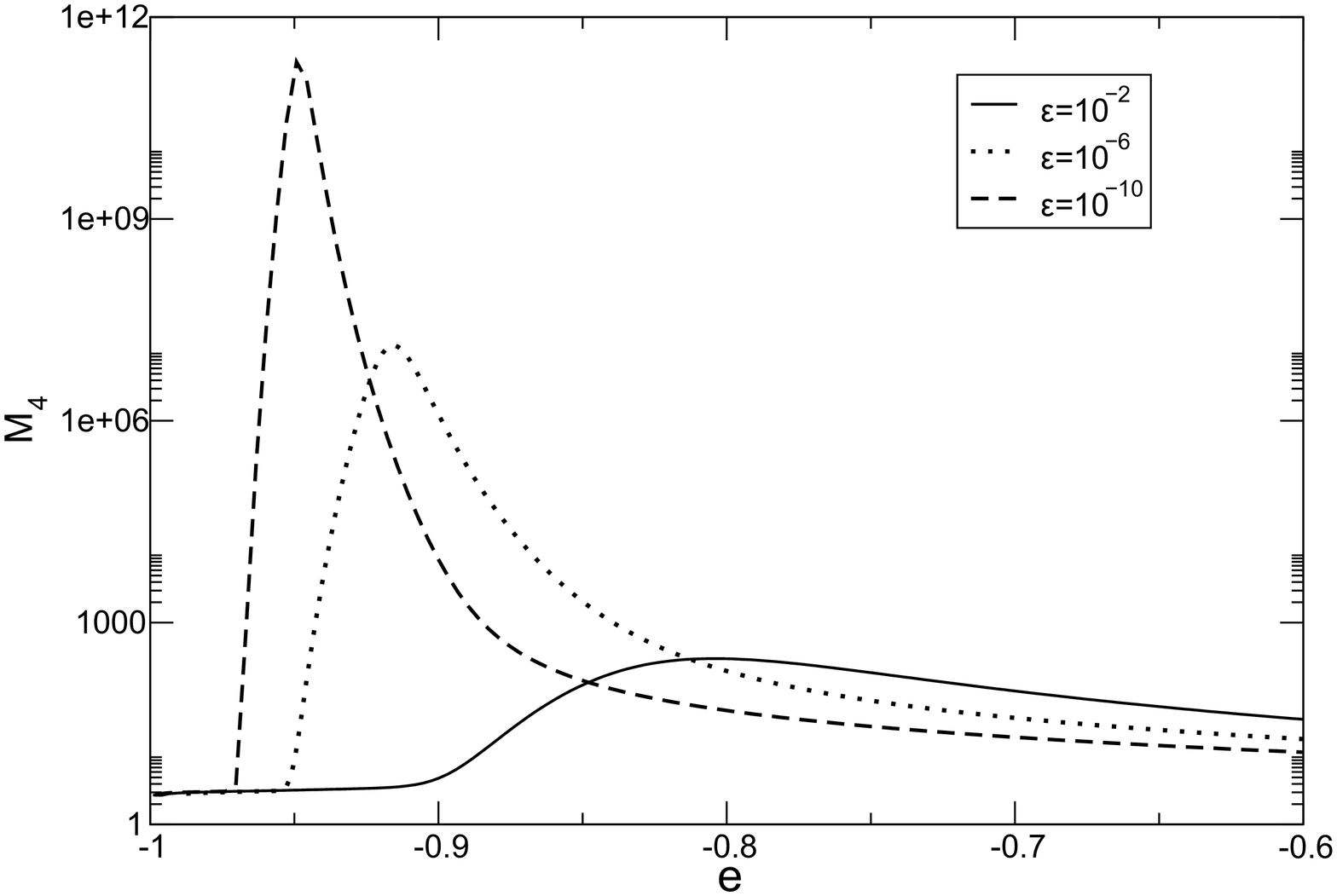}}}
\end{center}
\caption{Left panel: Standard deviation $\sigma$ as a function of the energy $e$ for a few values of $\epsilon$.
Right panel: Fourth moment $M_4$ (kurtosis).}
\label{fig2}
\end{figure}

\begin{figure}[ptb]
\begin{center}
\scalebox{0.3}{{\includegraphics{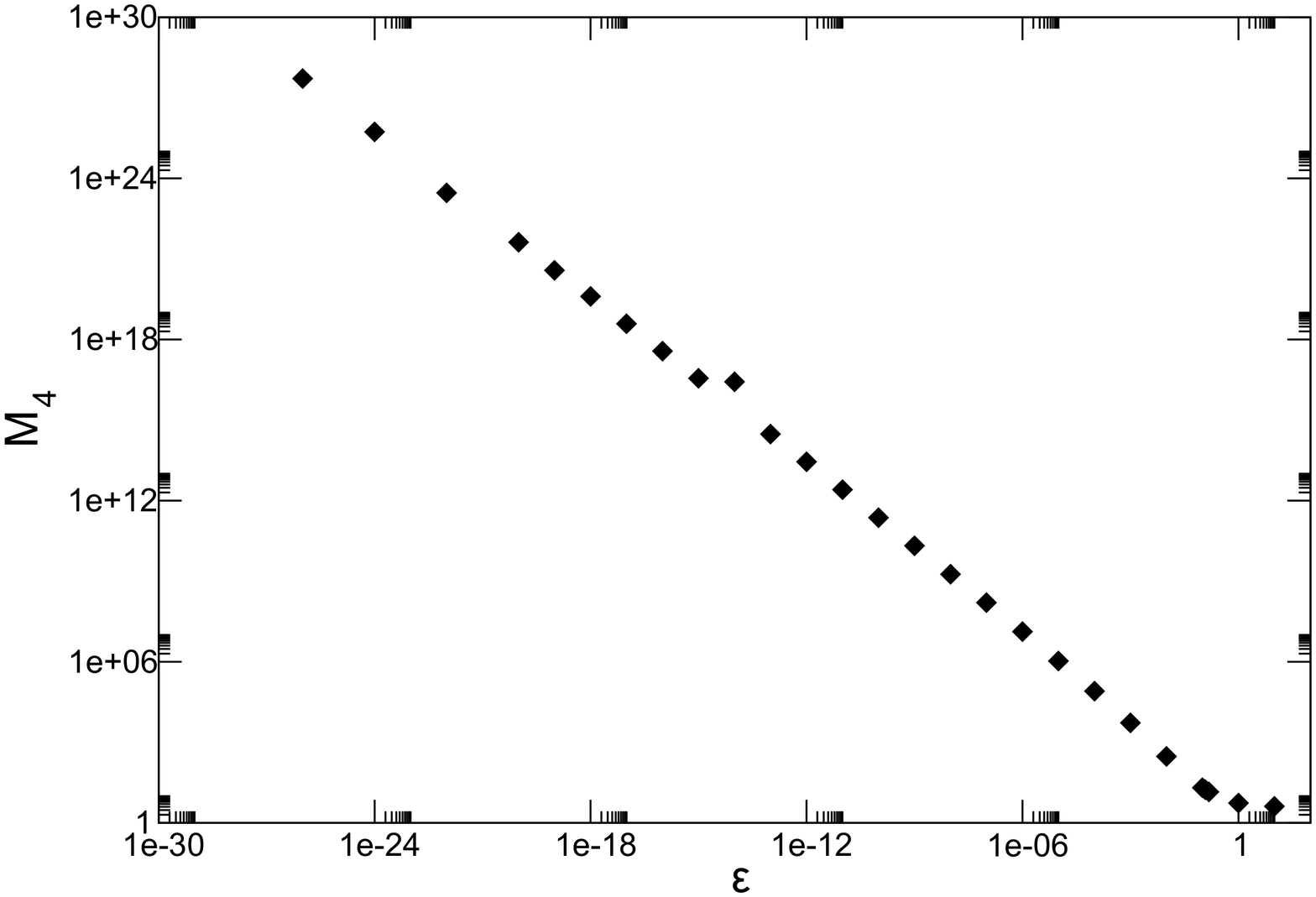}}}
\scalebox{0.3}{{\includegraphics{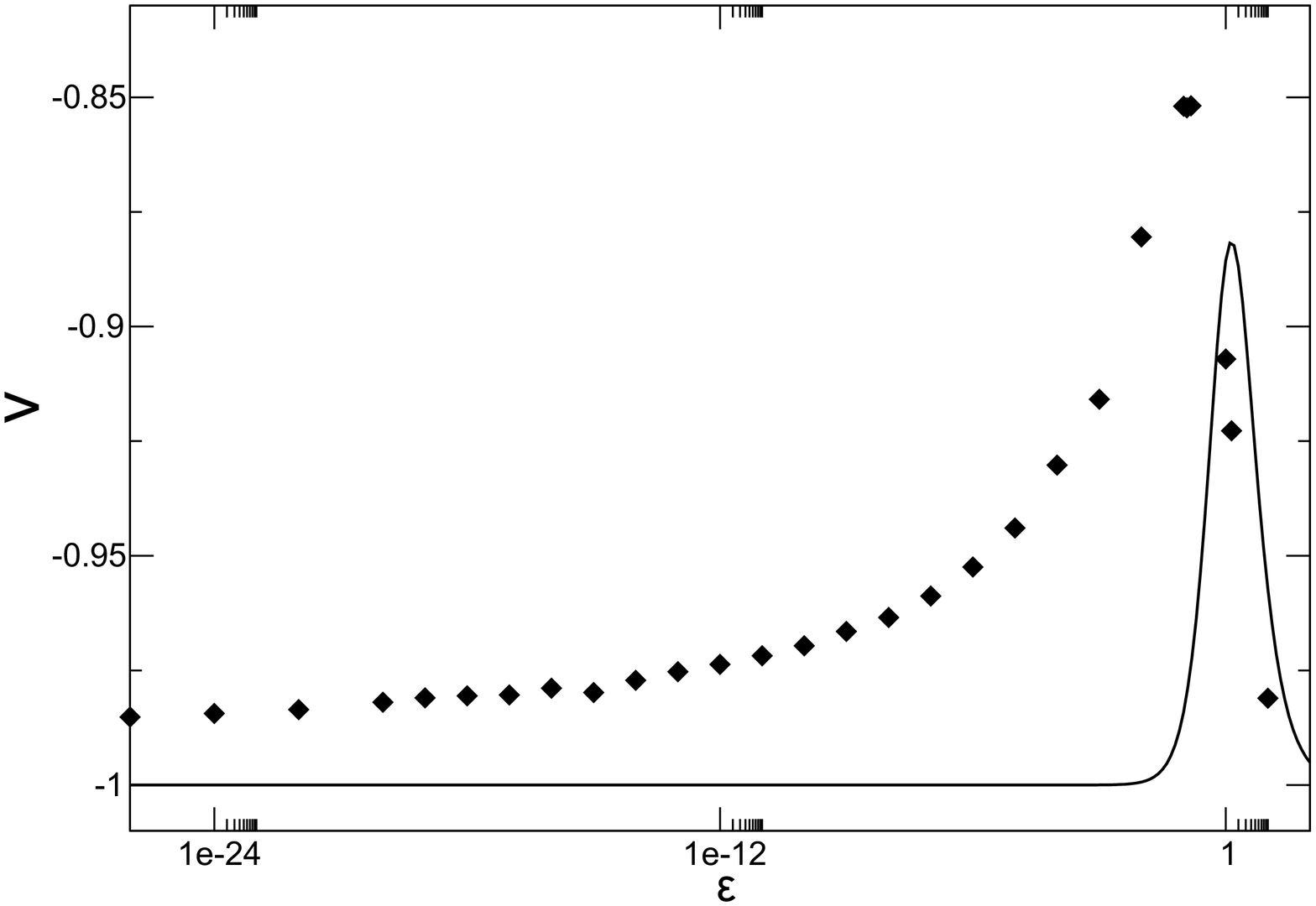}}}
\end{center}
\caption{Left panel: Maximum vale of $M_4$ as a function of $\epsilon$.
Right panel: Value of the potential per particle at the Kurtosis maximum (diamonds) and expected phase
transition from eq.~(\ref{vctrans}) (continuous line) as a function of $\epsilon$.}
\label{fig3}
\end{figure}

A more interesting feature of this model, also shared by the class of simplified models in next section, is visible
if one considers the first moments of the spatial distribution $\rho(\theta)$. Since the latter is an even function
all odd moments vanish. We consider here the dispersion $\sigma$ of the distribution
\begin{equation}
\sigma^2=\langle\theta^2\rangle=\int_{-\pi}^{\pi} d\theta\,\theta^2\rho(\theta),
\label{sigmadef}
\end{equation}
and the fourth momentum (kurtosis) $M_4$ of the reduced variable $\phi\equiv\theta/\sigma$:
\begin{equation}
M_4=\langle\phi^4\rangle=\frac{1}{\sigma^4}\int_{-\pi}^{\pi} d\theta\,\theta^4\rho(\theta).
\label{kurtdef}
\end{equation}
The kurtosis is very sensitive to minor changes in the distribution function, and any phase transition should come about
as a discontinuity in $M_4$ as a function of energy, or its derivative (and $\sigma$ also).
Figure~\ref{fig2} shows $\sigma$ and $M_4$ as a function of energy obtained using the EM method.
These results were tested for numerical convergence with a maximum relative error in the entropy of $10^{-8}$ at each energy value.
For small values of $\epsilon$ a very strong maximum for $M_4$ is obtained alongside a fast
variation of the derivative of $\sigma$.
This corresponds to
a change of regime in the distribution function, from a core-halo structure to a core only regime,
in the sense that in the latter case the halo is so extremely diffuse that it can be considered as nonexistent for any practical purposes.
The distribution in the halo cannot vanish for any coordinate if the mean-field self-consistent description is valid,
as it is the solution of a self-consistent equation of the form $\rho(\theta)\propto\exp(-\beta v(\theta))$, and therefore
this would imply a diverging positive potential.
The maximum of $M_4$ and the corresponding value of the potential per particle $v$ as a function of $\epsilon$
obtained using the EM method are shown in figure~\ref{fig3} for values from $\epsilon=10^{-26}$ up to $\epsilon=10$.
From our results it seems that a genuine phase transition only occurs for
$\epsilon\rightarrow 0$, corresponding to a collapse of the system into a single point for $T=0$.
The corresponding values of the predicted phase transition in eq.~(\ref{vctrans}) are also shown in the right panel of fig.~\ref{fig3} and exhibits a similar
qualitative behaviour as the curve corresponding to the actual regime change. Even though the presence of a critical point
satisfying the KSS criterion may somehow induce this change of regime,
we will show below that simplified models with no related critical points can exhibit the same type of behaviour.

\section{A class of simplified models}

In this section we discuss a class of solvable models with long range interactions, exhibiting
a behaviour in many aspects similar to the ring model, with a first order phase transition from a quasi-homogeneous
to a strongly inhomogeneous states, despite the fact that its potential energy has no critical (saddle) points.
Let us consider a pair interaction potential of the form:
\begin{equation}
V_{ij}=-u(r_i)u(r_j),
\label{solpot}
\end{equation}
with $u(r)$ a smooth positive function of the distance to the origin of the coordinate system such
that $V_{ij}$ is attractive, and the Hamiltonian
given by eq.~(\ref{hamgen}). The number of dimensions is restricted here to one, but similar solvable models can be obtained
for any dimensions following the same procedure.
The class of models corresponding to eq.~(\ref{solpot}) comprises both the Thirring (with a Kac $1/N$ regularization factor) and HMF
(at equilibrium) models as particular cases~\cite{thirring,antoni}. This whole class of models is solvable in the limit $N\rightarrow\infty$ (see below).
The mean field description is exact in this limit and the $N$-particle equilibrium distribution
function $f_N$ fully factorizes as~\cite{chavanis2}:
\begin{equation}
f_N(p_1,r_1,\ldots,p_N,r_N)=\prod_{i=1}^N f_1(p_i,r_i)+{\cal O}(1/N),
\label{fnfact}
\end{equation}
where $f_1$ is the one particle (equilibrium) distribution function.
The equilibrium properties can be computed using large deviation techniques~\cite{barre,ellis}, but here we chose to use
the approach in~\cite{metodo} which is more intuitive for the present purposes. The equilibrium properties of the system
in the microcanonical ensemble can be determined by maximizing either the Boltzmann or Gibbs entropy~\cite{reichl}, respectively:
\begin{equation}
S_B(E)=\log\left[\int\delta(H(p,q)-E)\:d\Gamma\right],
\label{boltzent}
\end{equation}
\begin{equation}
S_G(E)=-\int f_N\ln f_N \:d\Gamma,
\label{gibbsent}
\end{equation}
where $H(p,q)$ is the $N$-particle Hamiltonian with canonical variables collectively denoted by $p,q$,
$d\Gamma$ a volume element in phase space and the Boltzmann constant is set to unity.
The Gibbs entropy is more appropriate in the present context as it will become clear below.

For a fully factorized distribution function, the Gibbs entropy per particle is written as
\begin{equation}
s(e)=-\int dp\:dr\, f_1(p,r)\ln f_1(p,r).
\label{opge}
\end{equation}
The equilibrium distribution is obtained by maximizing $s(e)$ with the normalization constraint:
\begin{equation}
\int dp\:dr\, f_1(p,r)=1,
\label{normconst}
\end{equation}
and the total energy constraint:
\begin{equation}
\int dp\:dr\, \frac{p^2}{2m}f_1(p,r)-\frac{1}{2} \int dp\:dr\:dp'\:dr'\, u(r)u(r')f_1(p,r)f_1(p',r')=E,
\label{energconstr}
\end{equation}
with $m$ the particle mass. Local maxima and saddles of the entropy correspond to meta-stable
and unstable states, respectively. We therefore obtain:
\begin{equation}
\delta S=-\ln f_1-1-\lambda-\beta e(p,r),
\label{eqdist}
\end{equation}
with $\lambda$ and $\beta$ Lagrange multipliers corresponding to constrains~(\ref{normconst}) and~(\ref{energconstr})
respectively, the mean-field energy per particle:
\begin{equation}
e(p,r)=\frac{p^2}{2m}+U(r),
\label{oneparten}
\end{equation}
and the mean-field potential:
\begin{equation}
U(r)=\int dr'\,V(r,r')=-u(r)\int dp'\:dr'\, u(r')f_1(p',r')=-\alpha u(r),
\label{mfpot}
\end{equation}
with
\begin{equation}
\alpha=\int dp'\:dr'\, u(r')f_1(p',r').
\label{alphadef}
\end{equation}
Equation~(\ref{eqdist}) then implies that the one particle distribution function is the solution of the self-consistent equation:
\begin{equation}
f_1(p,r)=C\exp\left[\beta\left(\alpha\: u(r)-\frac{p^2}{2m}\right)\right],
\label{eqdist2}
\end{equation}
with $C=\exp(-\lambda-1)$ the normalization constant. Plugging eq.~(\ref{eqdist2}) into~(\ref{alphadef}) and integrating over the moments
we finally obtain:
\begin{equation}
{\overline{\alpha}}/{\beta}=\int dr\,u(r)\:e^{\overline{\alpha}\:u(r)} \times\left[\int dr\,e^{\overline{\alpha}\:u(r)} \right]^{-1},
\label{algeq}
\end{equation}
with $\overline{\alpha}\equiv\beta\alpha$. Equation~(\ref{algeq}) is a transcendental equation for $\alpha$ (or $\overline\alpha$),
and the equilibrium distribution function is then obtained from eq.~(\ref{eqdist2}).
The stability of this solution must also be determined, and may depend on the ensemble considered if they are not equivalent~\cite{barre,metodo},
i.~e.\ we must impose that $f_1$ maximizes the entropy or minimizes the free energy, for the microcanonical and canonical ensemble,
respectively. Different models are obtained by specifying the physical space range and the function $u(r)$.

Let us consider the function in one-dimension:
\begin{equation}
u(r)=\frac{\sqrt{2\epsilon}}{\sqrt{r+\epsilon}},
\label{exu}
\end{equation}
with $r=|x|$, $\epsilon>0$ a softening parameter and $x$ the spatial coordinate.
The system is enclosed in a one-dimensional box defined by $L_1<x<L_2$ and $m=1$. The numerator in the right-hand
side of eq.~(\ref{exu}) is chosen such that the minimal energy per particle is $-1$.
In order to straightforwardly apply Morse theory the potential must be a Morse function,
and therefore it must be a smooth (analytical) function in all configuration space.
This is ensured by taking $L_1=0$ and $L_2=L>0$. Even though the resulting model is quite artificial,
it still allows to discuss the main results of this section. The algebraic equation~(\ref{algeq}) is here:
\begin{equation}
{\overline{\alpha}}/{\beta}=Q(\overline{\alpha}),
\label{algeqex}
\end{equation}
with
\begin{eqnarray}
\lefteqn{Q(\overline{\alpha})=\left[4\epsilon{\rm Ei}\left(-A\overline{\alpha}\right)-4\epsilon{\rm Ei}\left(-\sqrt{2}\:\overline{\alpha}\right)
+2\sqrt{2\epsilon}\sqrt{L+\epsilon}\:e^{A\overline{\alpha}}-2\epsilon\sqrt{2}e^{\sqrt{2}\:\overline{\alpha}}\right]}
\nonumber\\
 & &\times\left\{(L+\epsilon)e^{A\overline{\alpha}}-\epsilon e^{\sqrt{2}\:\overline{\alpha}}\left(\sqrt{2}\:\overline{\alpha}+1\right)
+2\epsilon\overline{\alpha}^2\left[{\rm Ei}\left(A\overline{\alpha}\right)-{\rm Ei}\left(-\sqrt{2}\:\overline{\alpha}\right)\right]\right\}^{-1},
\label{delq}
\end{eqnarray}
where $\Ei$ is the exponential integral function and
$A={\sqrt{2\epsilon}}/{\sqrt{L+\epsilon}}$.
The energy per particle is obtained from eqs.~(\ref{oneparten}) and~(\ref{mfpot}) and the solution of eq.~(\ref{algeqex}):
\begin{equation}
e=\frac{1}{2\beta}-\frac{\alpha^2}{2}.
\label{mfenerg}
\end{equation}
In this expression the potential energy is not the average value of the mean-field potential, but half this value.

\begin{figure}[ptb]
\begin{center}
\scalebox{0.3}{{\includegraphics{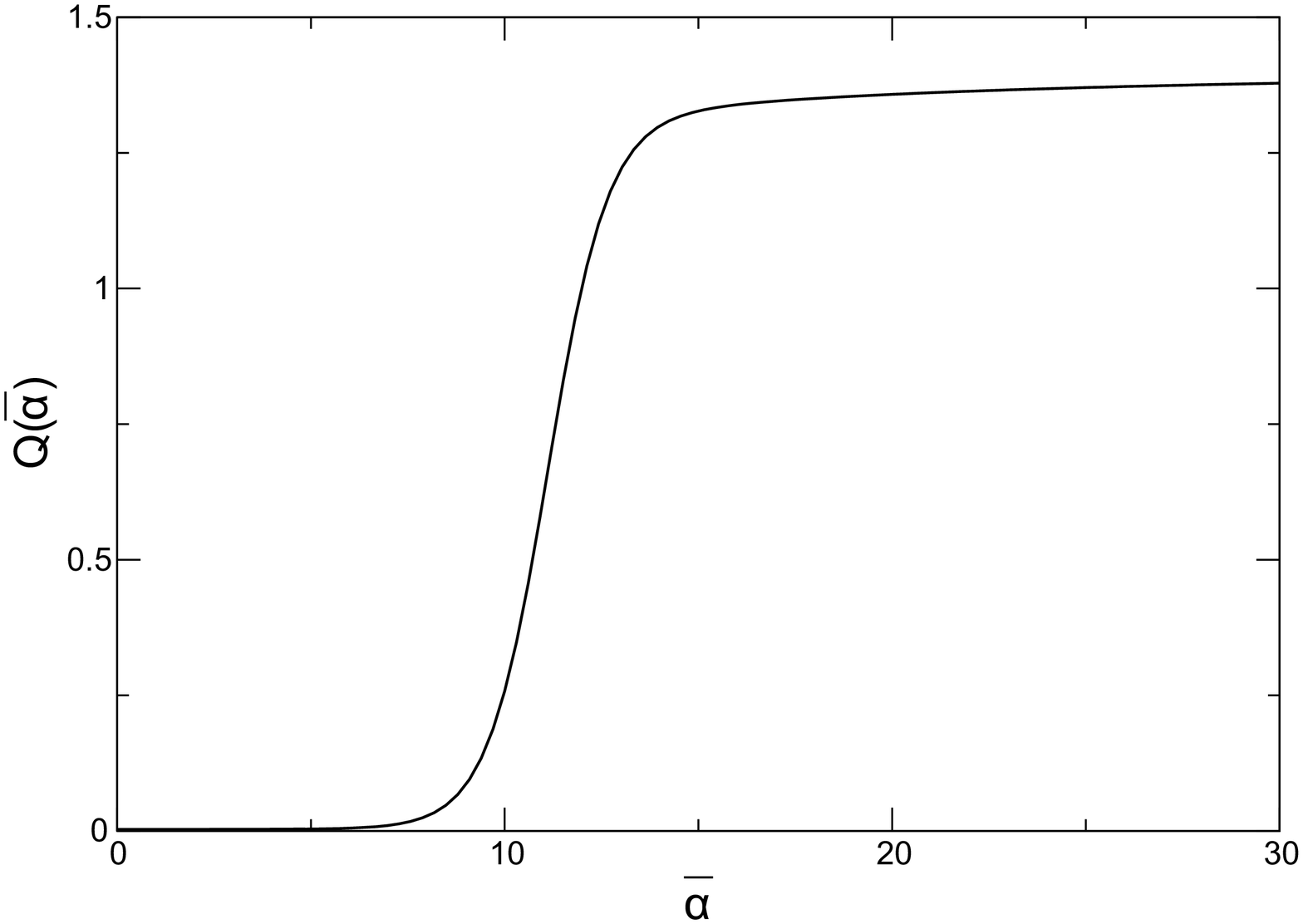}}}
\scalebox{0.3}{{\includegraphics{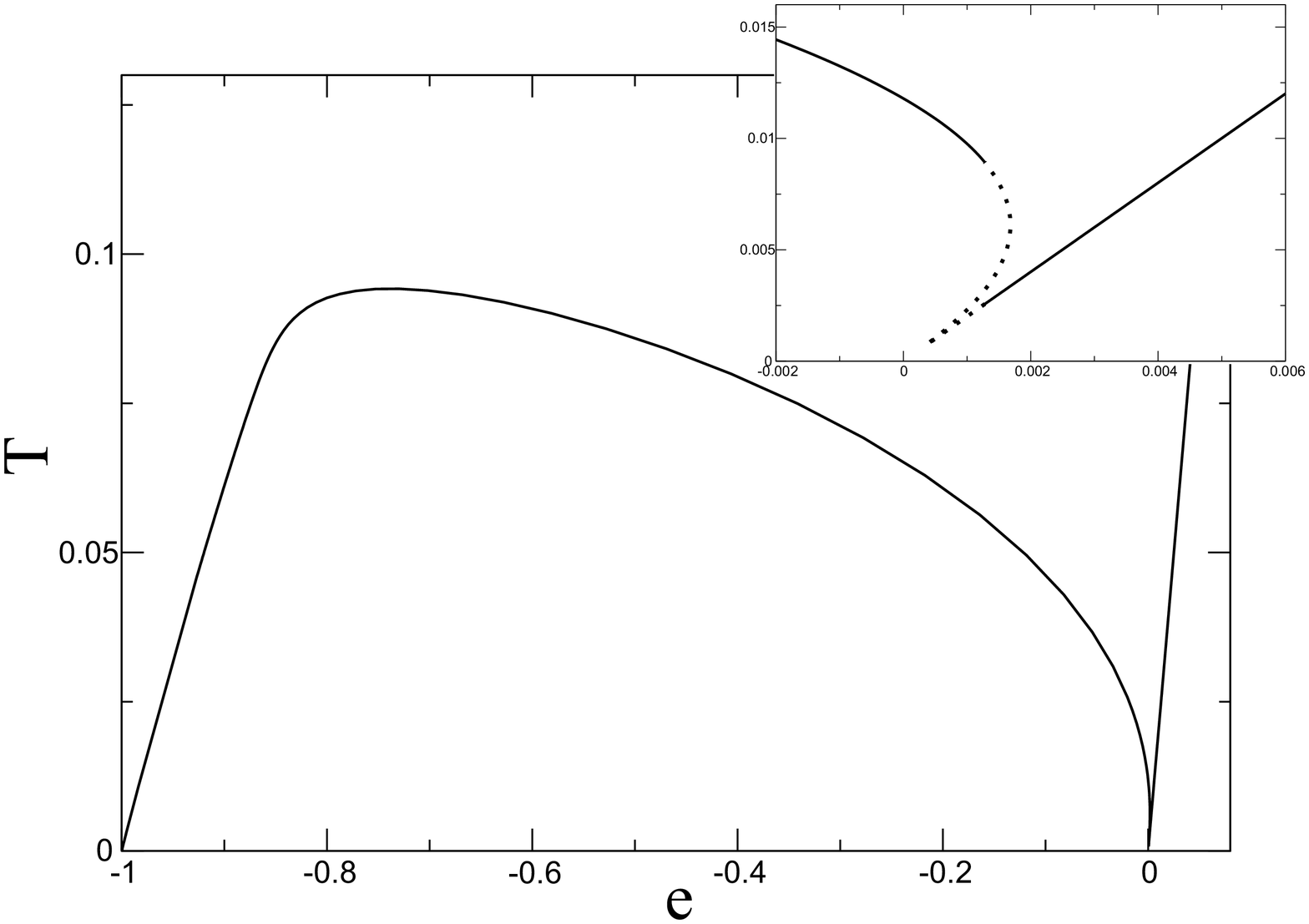}}}
\end{center}
\caption{Left panel: $Q(\overline{\alpha})$ for $\epsilon=10^{-6}$, with $Q(\overline{\alpha}=0)\approx0.028$.
Right panel: Caloric curve. The temperature jump is visible on the insert. The dotted line represents the meta-stable and unstable branches.}
\label{fig4}
\end{figure}

\begin{figure}[ptb]
\begin{center}
\scalebox{0.3}{{\includegraphics{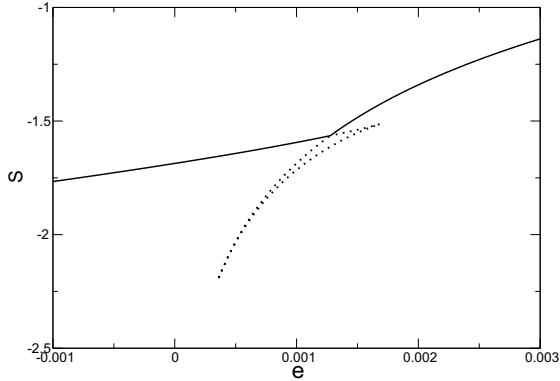}}}
\end{center}
\caption{Equilibrium entropy $S$ in the first order transition region. The dotted line shows the meta-stable and unstable branches.}
\label{fig3b}
\end{figure}

Without loss of generality we can choose $L=1$.
The graph of $Q(\overline{\alpha})$ for $\epsilon=10^{-6}$ is given in fig.~\ref{fig4}.
Equation~(\ref{algeqex}) has three different solutions for $10.6\lesssim\beta\lesssim1386.2$,
or equivalently $0.00072\lesssim T\lesssim0.094$ and $-0.738\lesssim e\lesssim0.00035$. Outside
this interval the algebraic equation~(\ref{algeqex}) has a single solution.
These different branches can be identified in the caloric curve shown in fig.~\ref{fig4},
where there are either one or three different energy values
for a given temperature, corresponding to the possible solutions of eq.~(\ref{algeqex}).
From figure~\ref{fig3b} we see that a first order phase transition occurs 
at $e\approx0.00128$ from an almost homogeneous state with positive heat capacity to a strongly inhomogeneous state with negative
heat capacity. The heat capacity changes sign at a lower energy.
A measure of the degree of inhomogeneity in the high energy phase is given by $Q(\overline{\alpha}=0)$,
which at its turn is a monotonously increasing function of $\epsilon$.
In the canonical ensemble the phase transition occurs with an energy jump~\cite{prrev}.
This behaviour is qualitatively the same for all small values of $\epsilon$. For $\epsilon\gtrsim 1.05\times 10^{-4}$ there is no phase transition.

\begin{figure}[ptb]
\begin{center}
\scalebox{0.3}{{\includegraphics{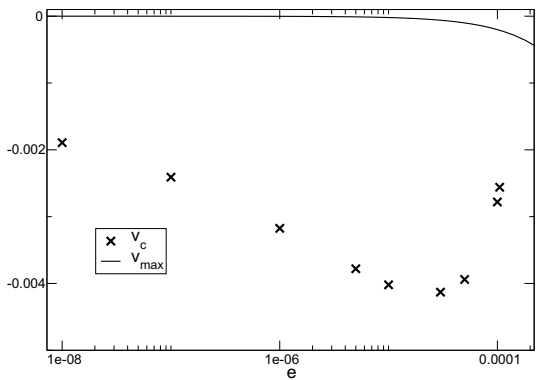}}}
\end{center}
\caption{Potential per particle $v_c$ at the phase transition for some values of $\epsilon$ obtained from the EM method, and the maximum value
of the potential per particle $v_{max}$.}
\label{fig7}
\end{figure}

The phase transition do not correspond to any stationary point in the potential of the system, as
it has no stationary points. The only topology change of ${\cal M}_v$ occurs at the potential minimum corresponding to $v=-1$, as for $v<-1$ the set
set ${\cal M}_v$ is empty. This topology change is not associated to any finite temperature phase transition. On the other hand, the manifold $\Sigma_v$ changes topology at $v=-1$
and $v=\epsilon/(L+\epsilon)\equiv v_{max}$, the latter corresponding to the maximum of the potential, the set $\Sigma_v$ being non-empty only for
$-1\leq v\leq v_{max}$. It is noteworthy that the phase transition in this model, as for the ring and HMF models, are a consequence of the potential being
bounded from above. Therefore one might expect that the phase transition is triggered by the topology change of $\Sigma_v$ at $v=v_{max}$.
Nevertheless the critical value $v_c$ of the potential at the transition is always smaller that $v_{max}$ as shown in fig.~\ref{fig7}.
Therefore, no topology change of either ${\cal M}_v$ or $\Sigma_v$ can be associated to the phase transition observed in this model.
A similar behaviour was also obtained for the spherical model~\cite{risau,angelani}.

\begin{figure}[ptb]
\begin{center}
\scalebox{0.45}{{\includegraphics{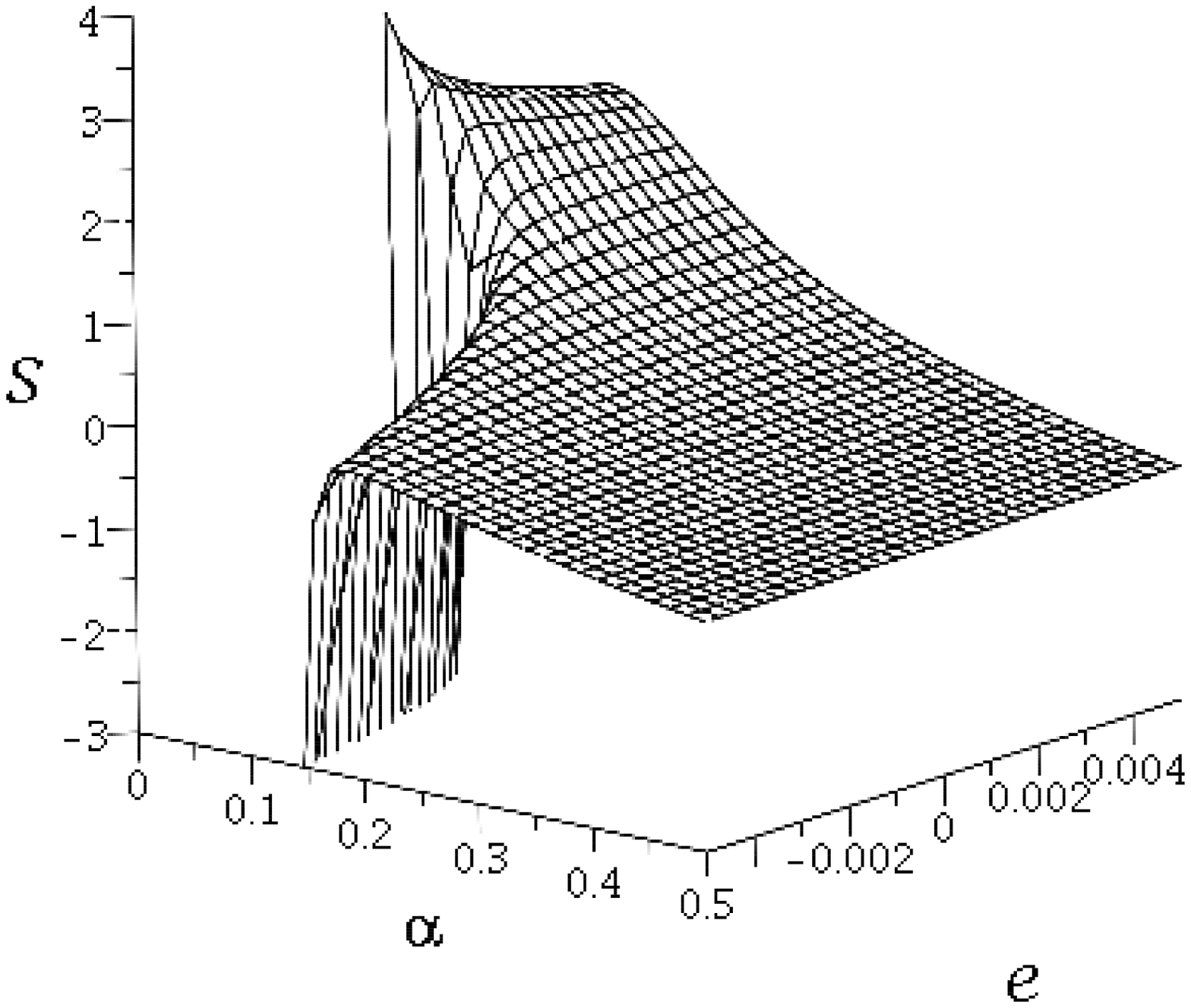}}}
\scalebox{0.3}{{\includegraphics{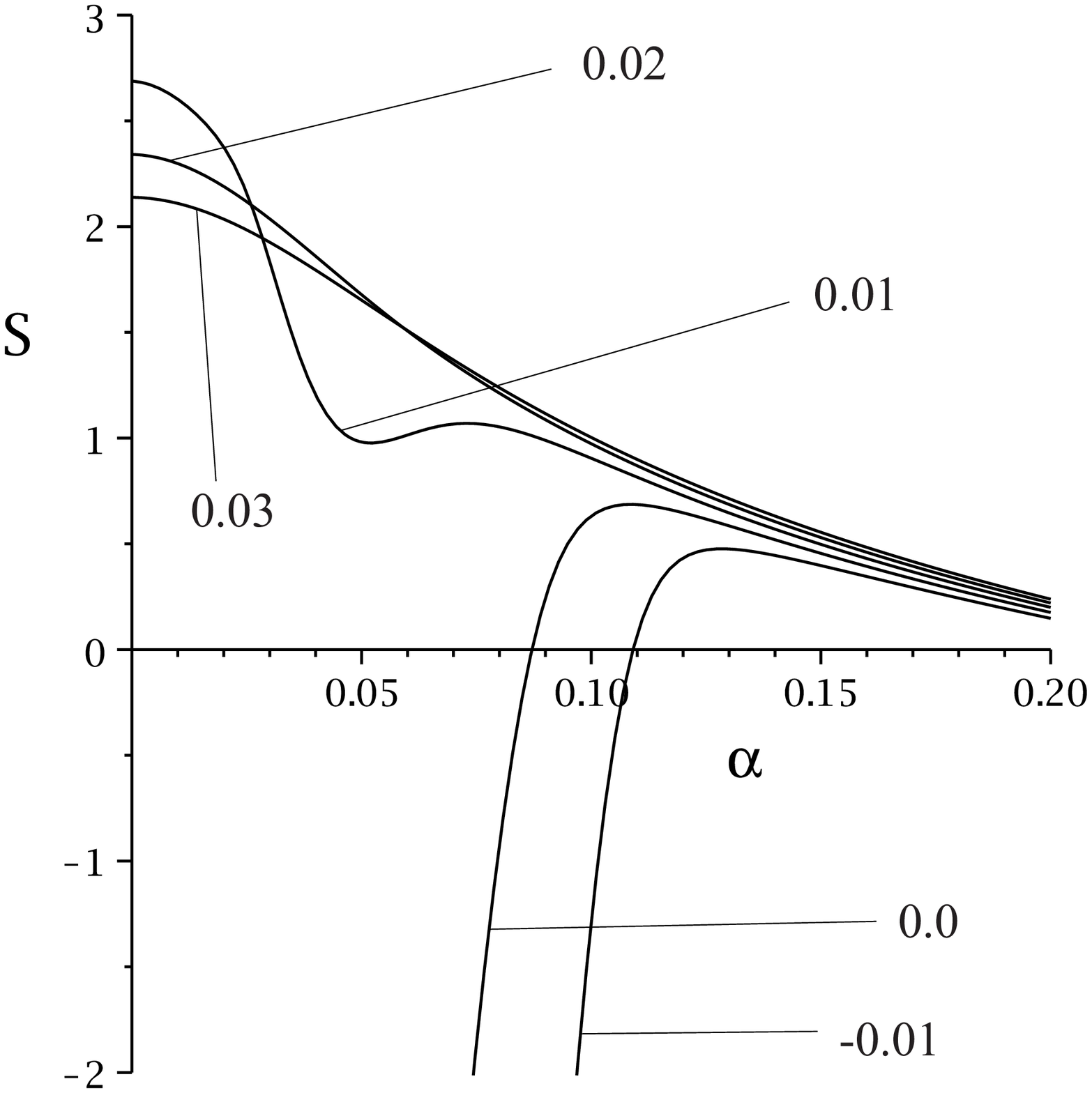}}}
\end{center}
\caption{Left panel: Entropy as a function of the energy $e$ and mean-field variable $\alpha$ on the region of the first order phase transition.
Right panel: Entropy as a function of $\alpha$ for different energy values. For some energy values a local maximum exists corresponding to
meta-stable states near the phase transition}
\label{fig3c}
\end{figure}

The origin of the phase transition can be explained by considering the entropy as a function of both the energy $e$ and the mean-field variable $\alpha$,
obtained from eq.~(\ref{gibbsent}) and the distribution function in eq.~(\ref{eqdist2}),
after eliminating $\beta$ from eq.~(\ref{mfenerg}), as:
\begin{equation}
s(e,\alpha)=2\beta e-\log C-\frac{1}{2}.
\label{entealpha}
\end{equation}
Note that the normalization factor $C$ in eq.~(\ref{eqdist2}) is also a function of $e$ and $\alpha$.
The function $s(e,\alpha)$ is shown on the left panel of figure~\ref{fig3c} in a region of the $(e,\alpha)$ containing
the first-order phase transition, where it is clear that it is not always a concave function for all values of its arguments.
The equilibrium entropy, as obtained above, is then given by:
\begin{equation}
s(e)={\rm max}_{\alpha}\:\: s(e,\alpha).
\label{semax}
\end{equation}
Even though $s(e,\alpha)$ is an analytical function of both its arguments,
the maximization in~(\ref{semax}) with respect to $\alpha$ yields a function with a non-analyticity at
the critical energy, as shown in fig.~(\ref{fig3b}).
Hahn and Kastner discussed this scenario for a system where the magnetization has a role similar to the variable $\alpha$~\cite{hahn1,hahn2}.
It is straightforward to see that provided $u(r)$ is a positive monotonously increasing function of $r$,
the qualitative behaviour of the resulting models is the same as the representative example just discussed above.

\begin{figure}[ptb]
\begin{center}
\scalebox{0.3}{{\includegraphics{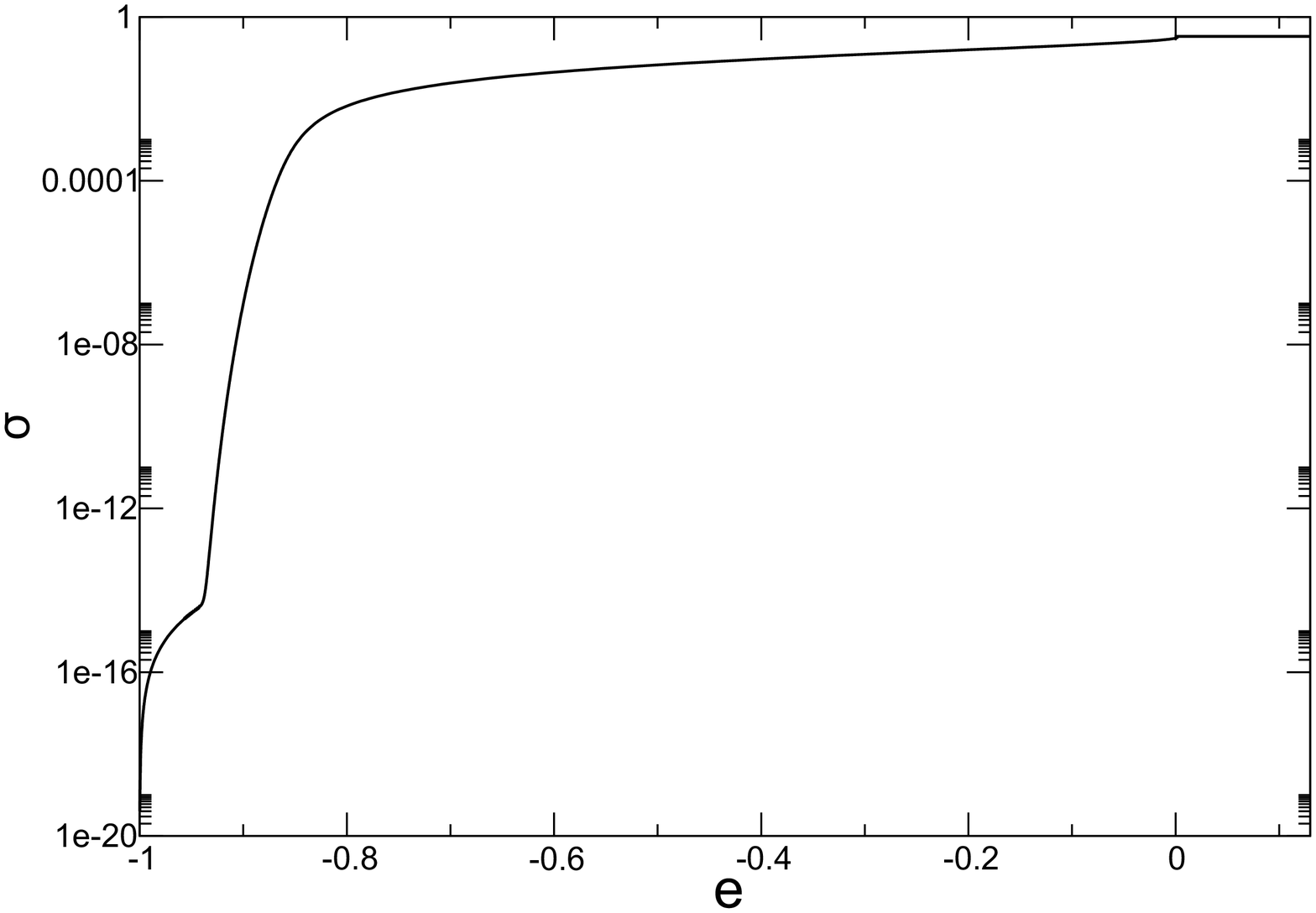}}}
\scalebox{0.3}{{\includegraphics{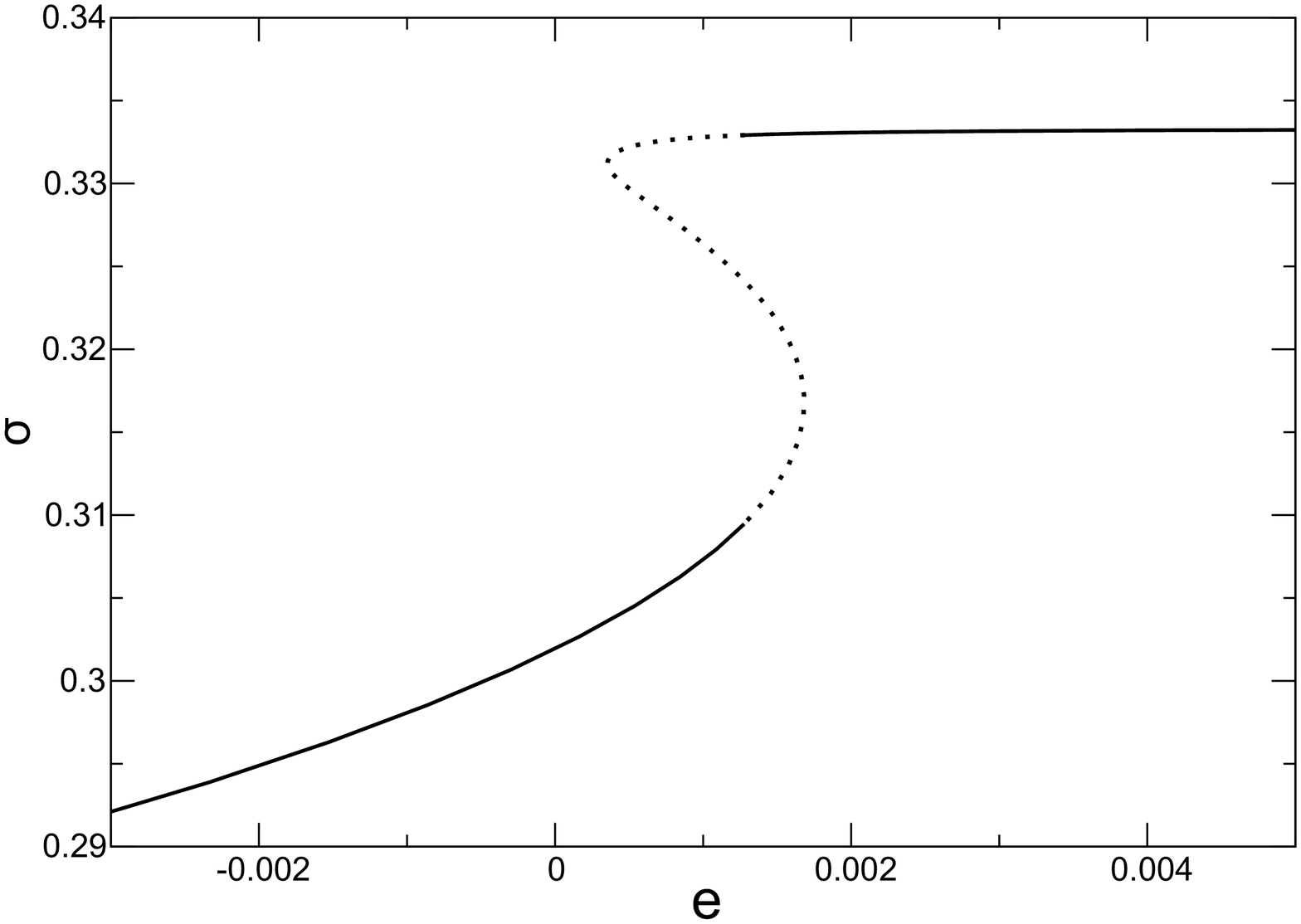}}}
\end{center}
\caption{Left panel: Mono-log plot of the dispersion $\sigma$ as a function of energy for $\epsilon=10^{-6}$.
Right panel: Zoom over the first-order transition region showing the solution for the stable (continuous lines), meta-stable and unstable
branches (dotted line).}
\label{fig5}
\end{figure}

\begin{figure}[ptb]
\begin{center}
\scalebox{0.3}{{\includegraphics{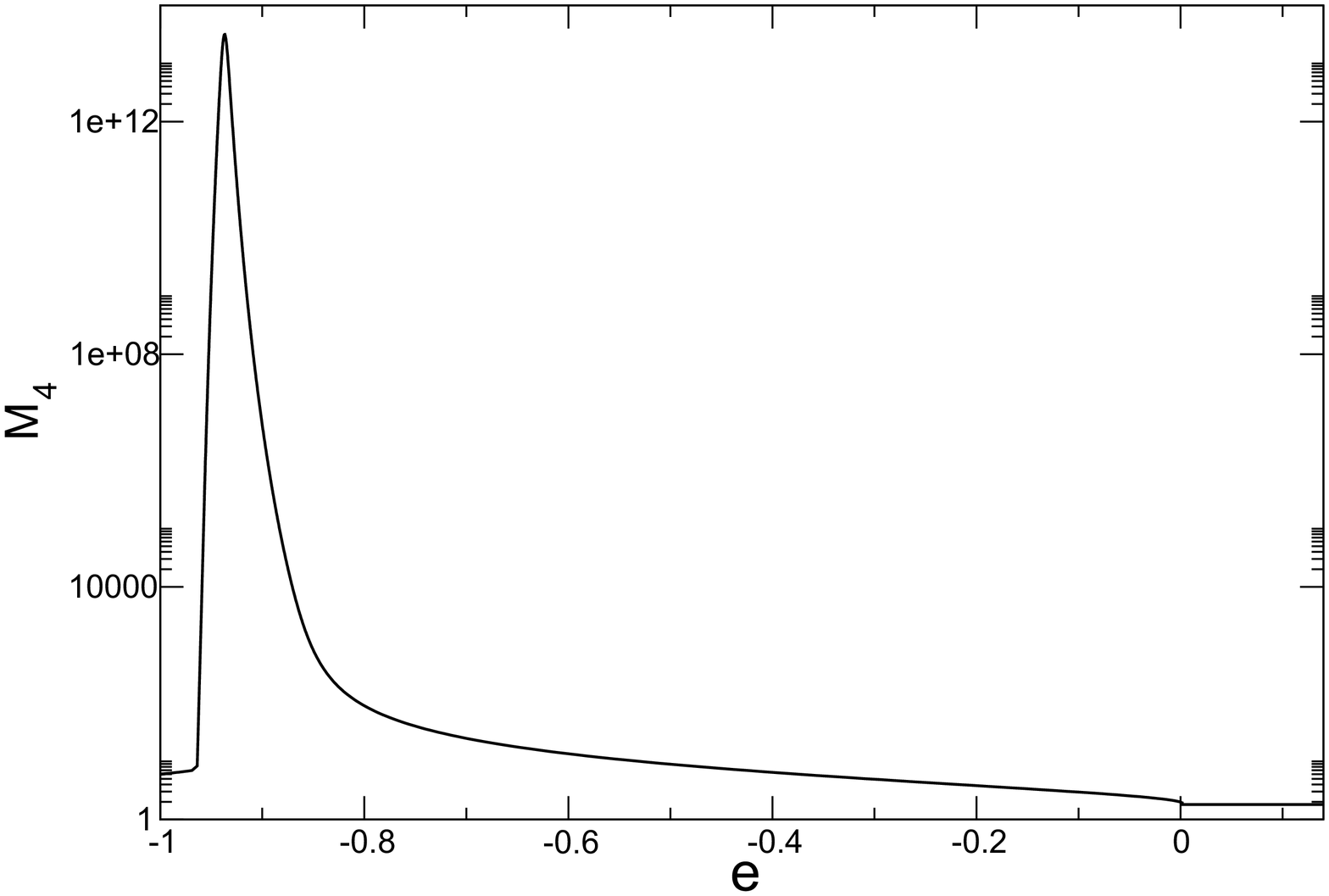}}}
\scalebox{0.3}{{\includegraphics{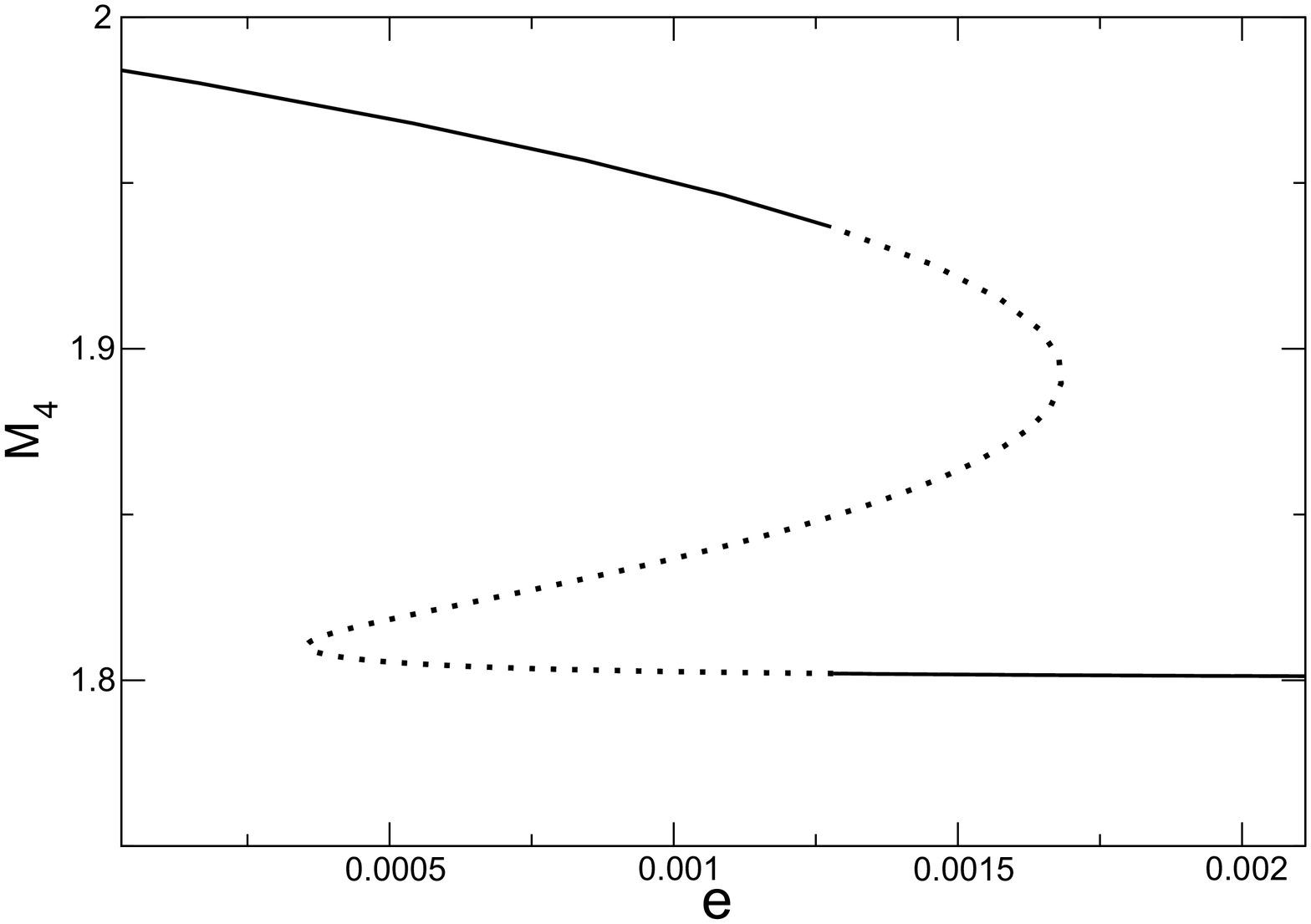}}}
\end{center}
\caption{Left panel: Fourth momentum $M_4$ of the reduced variable $\overline{r}=r/\sigma$.
Right panel: $M_4$ in the first order transition with the meta-stable and unstable branches given by the dotted line.}
\label{fig6}
\end{figure}

\section{Discussion and Concluding Remarks}

We have shown that the conjectured new phase transition for the ring model, as discussed in Ref.~\cite{ring3}
is not observed in our results, obtained from MC, MD and EM methods.
The same model nevertheless displays a marked change of regime from a core-halo to a core only structure.
For the core structure, the probability of a particle being in the halo
is so small that it can be considered as vanishing for any practical purposes. In the previous section we introduced a class
of simplified solvable models,
and studied in detail a representative case as defined by the choice of potential energy in eqs.~(\ref{solpot}) and~(\ref{exu}).
This particular model has properties very similar to those of the ring and Thirring models, and to the simplified model
in Ref.~\cite{casetti2}. It has a first order phase transition from an almost homogeneous phase to a clustered phase,
and a core-halo to core structural transition. The important point for this class of models, with a monotonous function $u(r)$,
is that it is possible to chose a potential with no stationary points, such that a first order
phase transition is obtained from the maximization of the entropy per particle $s(e,\alpha)$ with respect to the variable
$\alpha$ given in eq.~(\ref{alphadef}). Even with stationary points in the potential, the entropy is obtained by the maximization in
eq.~(\ref{semax}) of a smooth non strictly concave function.
This origin for phase transitions in long-range interacting systems was discussed
by Hahn and Kastner~\cite{kastner1,hahn1,hahn2}. In fact, it seems that the mean-field nature of long range interacting systems~\cite{chavanis2} is such
that this mechanism is ubiquitous in these systems, regardless of the existence of a topology change in phase space. As a perspective,
it would be interesting to investigate if a similar behaviour is also observed for two and three-dimensional self-gravitating systems,
as they also display a core-halo equilibrium structure~\cite{levin0}.

\section{Acknowledgments}

The authors would like to thank CNPq and CAPES (Brazil) for partial financial support.

\end{document}